\documentclass[ES]{rcftex}
\usepackage[english]{babel}

\titleEN{Black Holes and the 2020 Nobel Prize in Physics} 
\title{Agujeros negros y el Premio Nobel de F\'isica de 2020}

\author{Yuri Bonder\toaff{\emailto} \& Benito A. Ju\'arez-Aubry}

\affiliations{
Instituto de Ciencias Nucleares, Universidad Nacional Aut\'onoma 
de M\'exico\\ Apartado Postal 70-543, Ciudad de M\'exico, 04510, M\'exico. bonder@nucleares.unam.mx\emailto
}

\abstractES{El Premio Nobel de F\'isica 2020 distingui\'o dos proyectos de investigaci\'on sobre agujeros negros, que son una de las predicciones m\'as sorprendentes de la Relatividad General. El premio se dividi\'o en dos partes. La primera mitad fue otorgada a Roger Penrose y reconoce los teoremas de singularidad que garantizan que los agujeros negros, encontrados matem\'aticamente desde una etapa temprana del estudio de la Relatividad General, no son meras configuraciones gravitacionales curiosas con alta simetr\'ia, sino predicciones robustas de la teor\'ia. La segunda mitad se otorg\'o a Andrea Ghez y a Reinhard Genzel quienes lideraron dos grupos independientes que realizaron complicadas observaciones del centro de nuestra galaxia, las cuales sugieren que ah\'i hay un agujero negro supermasivo. En esta nota se explican las ideas principales de la teor\'ia de la relatividad general, as\'i como las caracter\'isticas principales de los agujeros negros. Se describen las dos obras premiadas en el mencionado premio Nobel.}

\abstractEN{The 2020 Nobel Prize in Physics distinguished two research projects on black holes, which are one of the most striking predictions of General Relativity. The prize was divided in two parts. The first half was awarded to Roger Penrose in recognition of his singularity theorems that guarantee that black holes, which were mathematically found since an early stage of the study of General Relativity, are not mere highly-symmetric, curious gravitational configurations, but robust predictions of the theory. The second half was awarded to Andrea Ghez and Reinhard Genzel who led two independent groups that carried out sophisticated observations of the center of our galaxy, which suggest that therein is a supermassive black hole. In this note, the main ideas of the theory of general relativity are briefly described, as well as the main features of black holes. The two works awarded in the aforementioned Nobel prize are described.}

\keyW{Nobel prize, general relativity, singularity theorems, black holes, supermassive black holes}

\begin{document}

\maketitle

\section{Introduction}

Gravity is currently best described by General Relativity (GR), a theory postulated by Albert Einstein in 1915. This theory successfully describes gravitational phenomena from mesoscopic to cosmological scales. Many experiments confirm GR predictions \cite{Will}: from the effects on Mercury's orbit and the bending of light rays, to the gravitational redshift that is nowadays detected in laboratories. More recently, an era began in which spacetime perturbations, known as gravitational waves, are regularly detected \cite{GW} and achieving high-precision cosmological measurements is no longer a dream \cite{PrecCosmo}. This ``golden age'' for GR has been crowned with many awards, including four Nobel prizes within the last decade.

This note concerns the 2020 Nobel Prize in Physics \cite{2020Nobel}, which recognises studies on one of the most striking GR predictions: the existence of black holes. Concretely, the above mentioned prize was awarded to Roger Penrose ``for the discovery that black hole formation is a robust prediction of the general theory of relativity'' and to Andrea Ghez and Reinhard Genzel ``for the discovery of a supermassive compact object at the centre of our galaxy,'' for which a black hole is the leading candidate.

GR states that gravity is an effect of spacetime geometry. The basic idea is that spacetime's geometry is ``deformed'' by matter (and all sorts of energy, including gravity itself), which, in turn, affects the matter propagation\footnote{For an accessible introduction to GR see, for example, Ref. \cite{GRDiv}.}. The basic variable describing this geometry is a pseudo-Riemannian metric tensor, which is governed by the Einstein field equations --the equations of GR--: a set of ten nonlinear, second-order, coupled differential equations. Importantly, the metric tensor allows one to compute curve lengths between any two spacetime points, thus filling spacetime with special curves that extremize such lengths. These curves are called \emph{geodesics}. For causally-related events, they represent the paths of light or of free point-like particles in the approximation where their effect on the spacetime curvature can be disregarded. The concept of geodesics plays a central role in the discoveries that have been awarded in the 2020 Nobel Prize, as well as in many other mathematical results and empirical observations.

While solving Einstein equations in full generality is extremely hard, only a few months after the publication of GR, the first exact solution was found by Karl Schwarzschild \cite{Schwarzschild} while he was serving in the German army during World War I. This solution describes the exterior of a spherical and nonrotating source, like a static star. More formally, the Schwarzschild metric is a solution of Einstein equations in vacuum for a static and spherically symmetric configuration. 

When the gravity source is extremely compact, the Schwarzschild solution has a series of mind-blowing features. If the source is confined to a radius smaller than the so-called Schwarzschild radius, $r_{\rm S}=2GM/c^2$, where $G$ and $c$ are Newton's gravitational constant and the speed of light in vacuum, respectively, and $M$ can be identified with the source's mass, then, the spacetime region contained inside $r_{\rm S}$ is causally disconnected from the exterior. That is, nothing can travel from the interior to the exterior. Since not even light, which achieves the maximal speed in the universe, can escape to the exterior, these objects are called black holes \cite{BHname}.

What is more striking is that any spherical and nonrotaing object that is compressed below its Schwarzschild radius must undergo a complete gravitational collapse. In other words, for such a body, no interaction can stop gravity from completely compressing the object. The outcome of this process was calculated \cite{Collapse} and the result was the formation of a spacetime region where the geometry is infinitely deformed, hidden behind an \emph{event horizon} from far-away observers. The event horizon defines the spacetime boundary separating the interior and exterior region of the black hole, and in this case sits precisely at the Schwarzschild radius. In fact, a black hole is \emph{defined} in mathematical terms by the existence of an event horizon. In any case, infinite spacetime deformation is a situation that lies outside the mathematical framework of GR, and such a pathological region is known as a \emph{singularity}.

Soon after, the generalisation of Schwarzschild's solution for electrically charged sources was found \cite{ReisnerNosrdstrom}, and in 1963, Roy Kerr discovered an exact solution of Einstein equations for rotating bodies \cite{Kerr} (stationary and axially symmetric solution). In all these generalisations there is an event horizon and a singularity hidden behind it.

Today, we know that stars that have a mass above a certain threshold undergo the above described gravitational collapse when their nuclear fuel, which generates outward pressure, comes to an end. This is because the dominant interaction at that stage is gravity, which is attractive. Still, in the mid 20th century it was hard to believe that such a process could actually occur in nature. At that time, two important questions remained open: Are the singularities a feature of the simplifying assumptions used to find exact solutions or are they generic consequences of the gravitational collapse? And, do black holes actually exist or are they mere mathematical curiosities of GR? Answers to these questions were worth the 2020 Nobel Prize in Physics.

\section{Singularity theorems}

As we have mentioned, the 2020 Nobel Prize in Physics was divided in two parts. The half that was awarded to Roger Penrose corresponds to mathematical studies on the emergence of spacetime singularities. Concretely, for the development of the first so-called \emph{singularity theorem} in 1965 \cite{Penrose}. Singularity theorems state that, under certain conditions, GR predicts the generation of a spacetime singularity. In particular, Ref. \cite{Penrose} shows that a singularity must develop inside a black hole.

The first task to study these theorems is to rigorously define singularities. Intuitively, a singularity is a spacetime region where the geometry has an infinite bending. However, this is not a good-enough definition in a theory where the dynamical variables describe the geometry itself. In addition, one needs to make sure that the variable describing the spacetime bending that blows up is coordinate independent as there are situations where problems in the coordinates give rise to infinities. The definition of a singularity that is used in these theorems reflects the fact that spacetime ``suddenly ends" in a singularity. Therefore, one can ``detect'' a singularity if there are geodesics that cannot be extended any further.

The second step in proving the singularity theorems concerns a set of geodesics, technically referred to as a congruence. A key equation for singularity theorems is one developed by Amal Kumar Raychaudhuri to characterise the congruence evolution \cite{Raych}. It is a consequence of Raychaidhuri's equation that if matter has nonnegative energy density --a reasonable physical assumption--, or more precisely, that certain \emph{energy conditions} hold \cite{EnergyCond}, then the congruences of causal geodesics tend to focus at some point within a finite geodesic parameter (e.g. within finite proper time for a test observer). Energy conditions imply geometrical conditions by the Einstein equations, and can thus be imposed even in spacetimes without matter.

The central concept introduced by Penrose to reach his conclusions is the following: consider now a two-dimensional, spacelike, closed set in spacetime. We can imagine such a set as a spherical surface or deformations thereof. There are two congruences of \emph{lightlike geodesics} --the paths of light rays-- orthogonal to this surface, defined by outgoing and ingoing sets of lightrays. Penrose defines such a two-dimensional set as a \emph{trapped surface} if both its ingoing and outgoing lightlike congruences have a focusing behaviour towards the surface itself. Thus, congruences tend to locally meet in the future: they are trapped! The key insight is that such trapped surfaces exist generically in the interior region of black hole spacetimes, whether they are static and spherically symmetric or not.

The Nobel-winning result that Penrose showed is that if a spacetime satisfies the Einstein equations and is globally hyperbolic (i.e., has well-posed dynamics) with a noncompact Cauchy surface, and a suitable energy condition holds, then the existence of a trapped surface implies the existence of a singularity -- the spacetime cannot be \emph{null geodesically complete} in the sense that there is at least one null ray that ``ends", thus defining a singularity.

The proof is by contradiction and, remarkably for its time, applies techniques of \emph{differential topology} to GR. It relies on defining the set of all points to the future of the Cauchy surface (technically in its future time development) that can be joined to the trapped surface by a future-leading smooth timelike curve. The boundary of this set is a compact lightlike surface, since lightrays meeting the trapped surface form a caustic in the future at finite geodesic parameter due to the energy condition. The contradiction is now achieved by assuming sufficiently ``long" null curves on the lightlike surface (which is acceptable if one has null geodesic completeness) and constructing a homeomorphism between this compact lightlike surface and the noncompact Cauchy surface of spacetime. Note that there are no assumptions regarding spacetime symmetries in this proof, and for the matter fields it is only assumed that they are physically realistic in the sense of the energy conditions.

The above mentioned result was the first singularity theorem to be proven, but not the last! Soon after, Roger Penrose and Stephen Hawking were able to generalise this result \cite{HawkingPenrose}. In particular, they were able to prove singularity theorems that do not require the assumption that spacetime must be globally hyperbolic and, importantly, used similar techniques to show that, under certain assumptions, the singularity at the beginning of the universe, i.e., the Big Bang, is an unavoidable consequence of GR.

\section{Astrophysical evidence of supermassive black holes}

According to the results discussed in the previous section, it is clear that the mathematical structure of GR predicts the existence of singularities, which are generally believed to lie inside a black hole. Yet, at the time it was unclear if there were black holes lurking out there in the universe. There are now several empirical indications of the existence of black holes; one of them was awarded with the other half of the 2020 Nobel Prize in Physics. Other experiments that have collected evidence supporting the existence of black holes include the detection of gravitational waves that are compatible with what is expected from binary black hole collisions \cite{GW}, and the ``picture'' of the object at the centre of the Messier 87 galaxy by the array called Event Horizon Telescope \cite{EHT}.

The black holes evidence that merited Andrea Ghez and Reinhard Genzel the 2020 Nobel Prize was obtained by observing the centre of our galaxy. While each one of the Nobel Laureates lead an independent experiment \cite{Exp1,Exp2}, both of these are similar and their basic principle is as follows: if one can trace the path of stars located close to the centre of the galaxy to the point that one can reconstruct their orbits, then, using Kepler's third law, it is possible to infer the mass of the central object. In addition, one can set an upper limit on the radius of the central object, since it must be smaller than the perihelion of the star's orbit. Interestingly, for the object at the centre of our galaxy, the mass is estimated to be several million times the mass of the Sun --the orbiting stars can be seen as test particles following geodesics--, but its size is such that it is smaller than the corresponding Schwarzschild radius. In addition, this central object does not seem to emit light. Therefore, the best suiting candidate to lie at the centre of our galaxy is a black hole.

It should be mentioned that these astronomical observations are daring, as there is dust in our light path to the centre of our galaxy, which absorbes visual light, and a high resolution is required. Therefore, the observations are done in the infrared and using sophisticated adaptive optics techniques to correct for atmospheric disturbances. The group of Andrea Ghez \cite{Exp1} uses a 10 m telescope at the W.M. Keck Observatory, which is located in Hawaii, while Reinhard Genzel's group \cite{Exp2} utilises the Very Large Telescope located in the Atacama desert, in the northern part of Chile.

Note that the object in the centre of our galaxy has a mass that is many orders of magnitude larger than the mass of a black hole that can be generated by the gravitational collapse of a star. Hence, this type of black holes are called \emph{supermassive black holes}. Remarkably, to date there is no compelling explanation of the mechanisms that produces these black holes \cite{Supermassive}. 

\section{Concluding remarks}

Black holes are one of the most intriguing predictions of GR and we now have mathematical and empirical evidence that these objects do exist in our universe. Crucially, the role of black holes in future scientific developments seems bright: on the one hand, understanding how supermassive black holes arise is one of the most important questions in astrophysics. On the other hand, singularities can be regarded as a breakdown of the theory at hand, GR, suggesting that there should be a new theory of gravity that resolves the singularities. It is believed that this new theory should also be compatible with the quantum description of the matter fields, and thus, it goes by the name of \emph{quantum gravity}.

In the quest for quantum gravity, black holes must play a central role. In fact, as Hawking postulated \cite{Hawking}, if quantum phenomena and gravity are considered together, black holes are not as dark as we think. They surprisingly emit radiation at the so-called \emph{Hawking temperature} and should, in principle, eventually evaporate. In any case, the yet-unknown quantum gravity theory will, very likely, revolutionise Physics, and black holes could be key players in the construction of such a theory. While the 2020 Nobel Prize was awarded to scientific projects that solved very important questions, they also raised new questions that should improve our understanding of concepts like space, time, and gravity.

\section{Acknowledgements}

Y. Bonder acknowledges support from the grant UNAM-DGAPA-PAPIIT IG100120. B.~A.~Ju\'arez-Aubry is supported by a CONACYT postdoctoral fellowship.



\end{document}